\documentclass[9pt,twocolumn,twoside]{osajnl}
\journal{ol} 
\setboolean{shortarticle}{true} 

\usepackage{textcomp}
\title{Tunable mid-infrared generation via wide-band four wave mixing in silicon nitride waveguides}

\author[1]{Abijith S. Kowligy}
\author[1]{Daniel D. Hickstein}
\author[1,2]{Alex Lind}
\author[1]{David R. Carlson}
\author[1]{Henry Timmers}
\author[3]{Nima Nader}
\author[1]{Daniel Maser}
\author[4]{Daron Westly}
\author[4]{Kartik Srinivasan}
\author[1,2]{Scott B. Papp}
\author[1,2]{Scott A. Diddams}

\affil[1]{Time and Frequency Division, NIST, Boulder, CO 80305 USA}
\affil[2]{Department of Physics, University of Colorado, Boulder, CO 80305 USA}
\affil[3]{Applied Physics Division, NIST, Boulder, CO 80305, USA}
\affil[4]{Center for Nanoscale Science and Technology, NIST, Gaithersburg, MD USA}

\affil[*]{Corresponding author: abijith.kowligy@gmail.com}

\begin{abstract}
We experimentally demonstrate wide-band (\textgreater100~THz) frequency down-conversion of near-infrared (NIR) femtosecond-scale pulses from an Er:fiber laser to the mid-infrared (MIR) using four-wave-mixing (FWM) in photonic-chip silicon-nitride waveguides. The engineered dispersion in the nanophotonic geometry, along with the wide transparency range of silicon nitride, enables large-detuning FWM phase-matching and results in tunable MIR from 2.6\textendash3.6~\textmu{}m on a single chip with 100-pJ-scale pump-pulse energies. Additionally, we observe up to 25~dB broadband parametric gain for NIR pulses when the FWM process is operated in a frequency up-conversion configuration. Our results demonstrate how integrated photonic circuits could realize multiple nonlinear optical phenomena on the same chip and lead to engineered synthesis of broadband, tunable, and coherent light across the NIR and MIR wavelength bands from fiber-based pumps. 
\end{abstract}
\setboolean{displaycopyright}{false}

\begin{document}
\maketitle

On-chip photonic platforms have allowed for exploration of novel classical and quantum nonlinear phenomena in highly controlled environments. This is due to the ability to precisely tailor the dispersion and nonlinearity for a given interaction \cite{obrien_photonic_2009,moss_new_2013}. For example, supercontinuum generation (SCG) can be tailored to provide multi-octave spectra \cite{hickstein_ultrabroadband_2017} and quantum frequency conversion for single photons can be engineered for large spectral translation \cite{li_efficient_2016}. Due to the availability of robust, inexpensive femtosecond laser sources in the near-infrared (NIR), a current goal for on-chip photonic devices is the conversion of light from the NIR to longer wavelengths in the mid-infrared (MIR). In particular, Er:fiber lasers are desirable pump sources due to their technological maturity and the surrounding telecom infrastructure. In this work, we utilize silicon nitride (Si$_3$N$_4$, henceforth SiN) nanophotonic waveguides to realize wide-band frequency conversion from the NIR to the MIR driven by Er:fiber lasers, and demonstrate tunability via dispersion engineering techniques. Such a compact source of MIR coherent light benefits many multi-disciplinary applications, including frequency comb generation and spectroscopy \cite{schliesser_mid-infrared_2012}. 

Nonlinear optical processes in second-order $\chi^{(2)}$ and third-order $\chi^{(3)}$ nonlinear media such as difference-frequency generation (DFG) \cite{erny_mid-infrared_2007,ruehl_widely-tunable_2012,krauth_broadly_2013,cruz_mid-infrared_2015} and SCG \cite{dudley_supercontinuum_2006,genty_fiber_2007,guo_mid-infrared_2018,grassani_highly_2018} have been explored for MIR generation. Conventionally, materials with strong $\chi^{(2)}$ nonlinearity have proved difficult to integrate with photonics platforms and often require quasi-phase-matching schemes that add complexity to the fabrication process \cite{bortz_quasi-phase-matched_1995,mayer_offset-free_2016}. In the case of SCG, the broad spectrum generated comes at a cost of spectral flux in any particular wavelength range. However, controlled frequency conversion that generates light in specific spectral regions is beneficial for many applications including spectroscopy and frequency referencing quantum cascade lasers \cite{chang_high_2016,cappelli_frequency_2016}.

\begin{figure}[t!]
\centering
\includegraphics[width=0.9\linewidth]{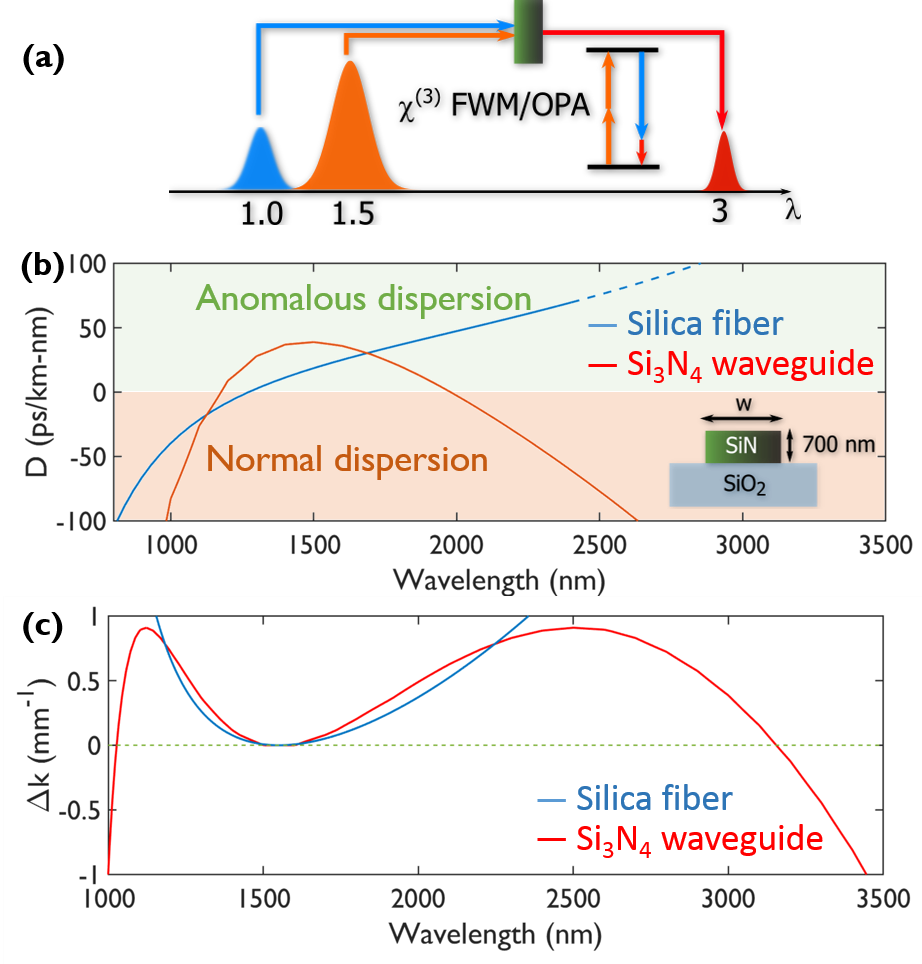}
\caption{(a) Frequency conversion from the NIR to the MIR via four-wave-mixing: two pump photons at 1.5 \textmu{}m mix with a 1.0 \textmu{}m signal photon to generate an idler at 3 \textmu{}m. (b) Compared to single-mode telecom-grade silica fibers, the dispersion in a nanophotonic waveguide (width $w = 3400$ nm) can be tailored over a broadband region due to the strong influence of the geometric dispersion, and lead to (c) phase-matching for wavelengths far from the pump.}
\label{fig:concept}
\end{figure}

Towards this goal, we utilize optical parametric amplification (OPA) using degenerate-pump FWM in $\chi^{(3)}$ materials ($\chi^{(3)}$-OPA) since phase-matching requires careful dispersion engineering and frequency conversion only occurs at desired spectral bands. In such a FWM process, two pump photons are annihilated, providing gain for a signal wavelength, and generate frequency converted light at the idler wavelength, where $2\omega_{p}=\omega_{s}+\omega_{i}$ is the energy conservation relation. The corresponding phase-matching condition is $\Delta\beta = 2\beta_{p}-\beta_{s}-\beta_{i}+2\gamma P = 0$, where $\beta$ denotes the wave-vector, $\gamma$ is the nonlinear parameter, and P is the pump peak power. Moreover, numerous materials with large $\chi^{(3)}$ nonlinearity can be processed with commonplace electron-beam- and photo-lithography techniques. This enables an integrated photonics realization of the $\chi^{(3)}$-OPA.

In previous work, such schemes have typically employed single-mode silica-based optical fibers due the favorable anomalous dispersion and low losses around 1.55 \textmu{}m \cite{agrawal_nonlinear_2012,radic_parametric_2008,marhic_fiber_2007}. Unfortunately, the tuning bandwidth is limited because the phase-mismatch grows quickly as the detuning from the pump increases (Fig.~\ref{fig:concept}(b)). To overcome this limitation, higher order dispersion in specialty highly-nonlinear fibers (HNLFs) and photonic-crystal fibers has been used to extend the phase-matching for large-detuning frequency conversion processes \cite{boggio_730-nm_2008,sharping_octave-spanning_2007}. However, the large absorption in conventional silica fibers beyond 2.4 \textmu{}m limits the generated wavelengths to the short-wave infrared (1.7--2.4 \textmu m). On the other hand, strong nonlinear losses at 1.5~\textmu{}m preclude the usage of Er:fiber technology in silicon- and chalcogenide-based devices and instead necessitate long-wavelength (\textgreater~2 \textmu{}m) pump lasers for FWM \cite{lamont_net-gain_2008,liu_mid-infrared_2010,kuyken_-chip_2011,kuyken_50_2011,lau_continuous-wave_2011,liu_bridging_2012,xing_mid-infrared_2017}, adding complexity to the system. 

Here, we exploit the wide transparency region (0.4\textendash{}4.5 \textmu{}m) and negligible two-photon absorption of SiN to realize phase-matched $\chi^{(3)}$-OPA pumped at 1.55 \textmu{}m to generate MIR light in the first atmospheric window (3--5 \textmu m) (Fig.~\ref{fig:concept}(c)). The sub-wavelength cross-sectional areas in the nanophotonic waveguide structures greatly enhance the nonlinearity ($\gamma \approx 1000$ W$^{-1}$ km$^{-1}$) and allow broadband dispersion engineering due to the increased influence of the geometric dispersion \cite{agrawal_nonlinear_2012}. Such dispersion-engineering techniques have allowed for broadband SCG in SiN waveguides \cite{carlson_self-referenced_2017} and octave-spanning Kerr frequency combs in the micro-resonator geometry \cite{spencer_optical-frequency_2018}. 

\begin{figure}[b!]
\centering
\includegraphics[width=\linewidth]{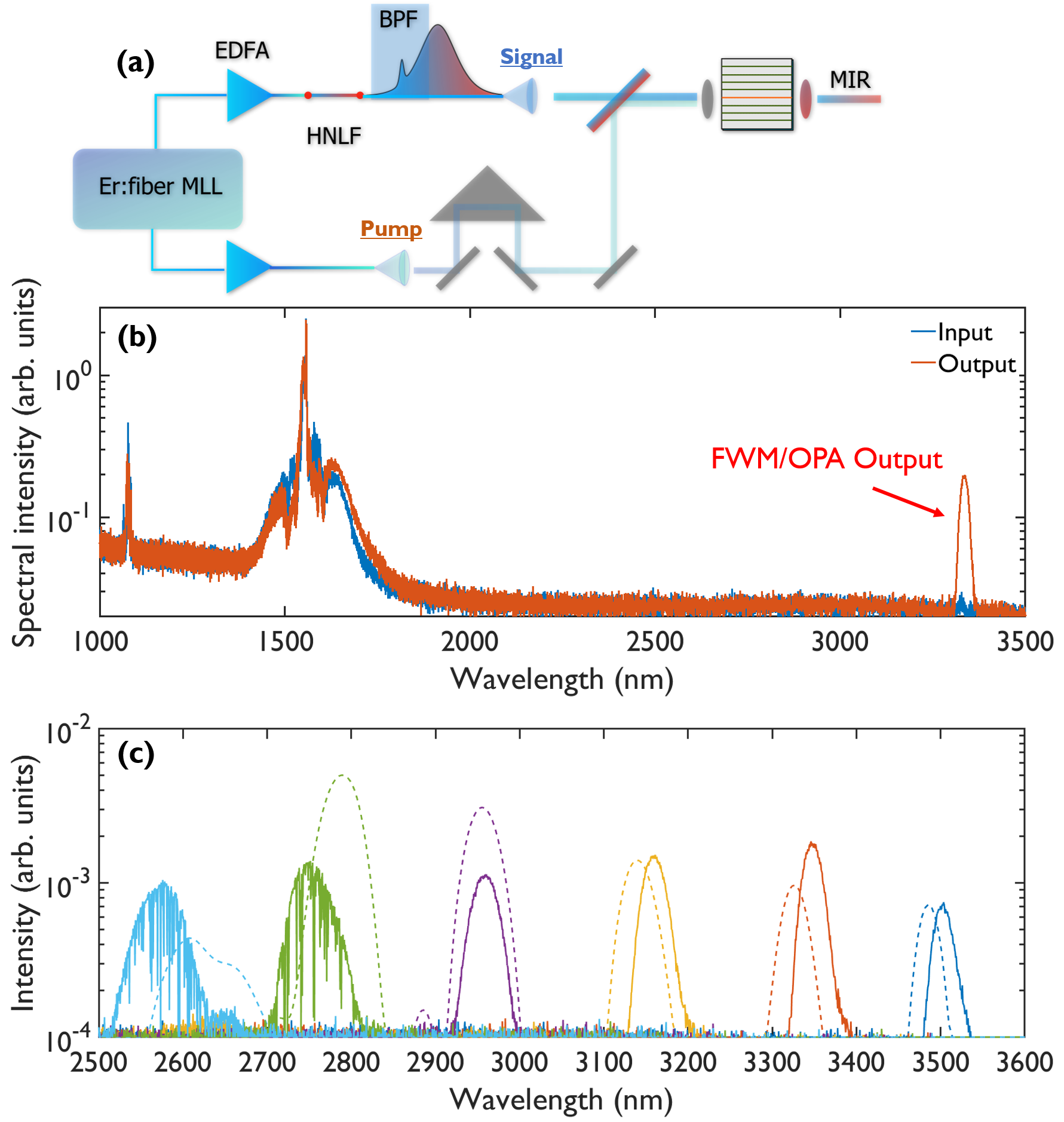}
\caption{(a) Two outputs from an Er:fiber laser are amplified using Er:doped fiber amplifiers (EDFA), and one amplified output is broadened in highly-nonlinear fiber (HNLF) to generate 1 \textmu{}m light. The amplified 1.5~\textmu{}m (pump) and 1.0 \textmu{}m (signal) beams are coupled into a SiN waveguide (inset) to generate MIR light via FWM. (b) When the pulses are temporally overlapped, the phase-matched idler is generated in the 3-\textmu{}m band (red). Without temporal overlap, the idler is not present (blue). (c) Tunable mid-infrared light is generated from 2.6--3.6 \textmu{}m for different waveguide widths in the experiment (solid lines) and from theoretical predictions (dashed lines).}
\label{fig:MIR}
\end{figure}

We generate tunable MIR light by using a single mode-locked Er:fiber laser (100~MHz repetition rate) to generate the pump pulses at 1.55 \textmu{}m (degenerate pump) and the signal at 1.07 \textmu{}m. In this configuration, the laser output is amplified in an erbium-doped fiber amplifier (EDFA) to yield a 50-fs, 1-nJ pump pulse. A second EDFA output is spectrally broadened in HNLF ($\gamma = 10$ W$^{-1}$ km$^{-1}$) and a bandpass filter around 1.0~\textmu m isolates the seed pulse (Fig.~\ref{fig:MIR}a). The two femtosecond-scale pulses are temporally overlapped using a tunable mechanical delay in the 1.5-\textmu m arm and spatially overlapped at a dichroic mirror before being coupled into the waveguide-chip using an aspheric lens. We note that this setup is analogous to the DFG-based MIR frequency comb generation in $\chi^{(2)}$ materials such as periodically poled lithium niobate \cite{cruz_mid-infrared_2015}, but accomplished using $\chi^{(3)}$ nonlinear media.

The stoichiometric SiN waveguides are fabricated via low-pressure chemical vapor deposition (LPCVD) with 700-nm thickness on a silicon dioxide substrate. Electron-beam lithography is used to define 1-cm-long ridge waveguides, which are air-clad except at the input and output facets where symmetric oxide cladding helps with coupling by making the waveguide modes symmetric. The waveguide input facet, in addition, has inverse taper couplers for efficient pump coupling (\textless~2~dB insertion loss) \cite{cardenas_high_2014}. The waveguide width at the output facet is not tapered, thereby limiting field overlap with the oxide cladding, for which absorption losses in the mid-IR are significant. On-chip pulse energies for the pump and seed are approximately 100 pJ and 50 pJ, respectively. The output is collected using a chalcogenide aspheric lens (0.58 NA) and measured using a free-space Fourier-transform spectrometer (FTS). 

\begin{figure}[b!]
\centering
\includegraphics[width=0.8\linewidth]{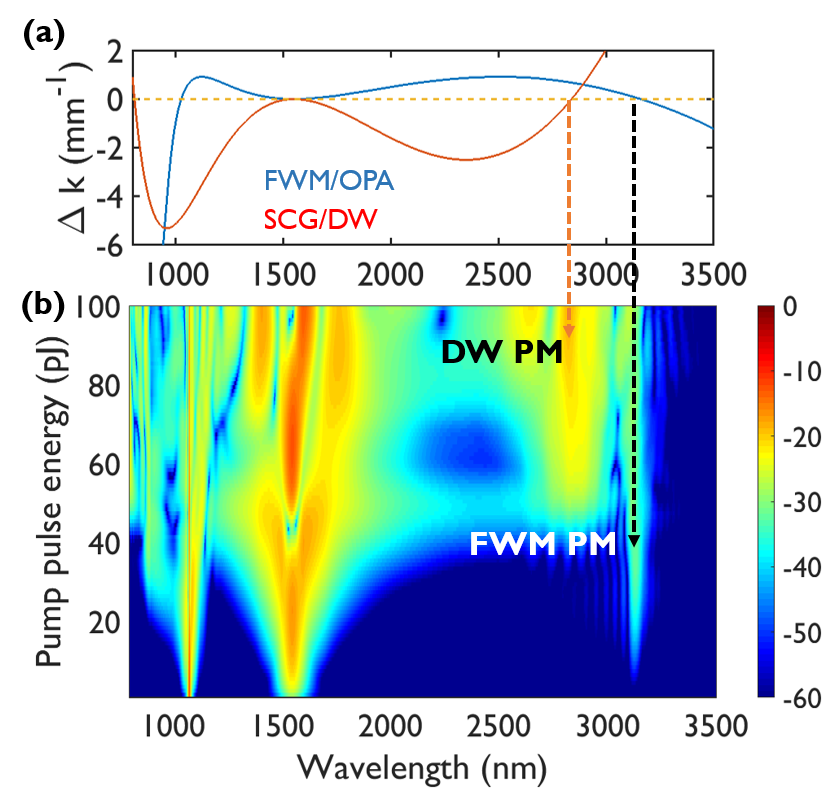}
\caption{(a) In silicon-nitride waveguides that support phase matching the FWM process (blue), SCG and associated dispersive-wave generation in the mid-infrared is also phase matched (red). This leads to parasitic soliton fission in the waveguide at high pump-pulse energies, and clamps the FWM conversion efficiency.
(b) Output spectra of a 10-mm long silicon-nitride waveguide with width w = 3400 nm as a function of pump-pulse energy. The signal energy is kept fixed at 10 pJ on chip. The arrows mark the predicted FWM and dispersive-wave phase-matched wavelengths. 
}
\label{fig:MIRSim}
\end{figure}

The waveguides with widths 3000--4000 nm have anomalous dispersion at 1.5 \textmu{}m (Fig.~\ref{fig:concept}(b)) and support phase matching for the FWM interaction. When the pump and signal pulses are temporally overlapped in the photonic waveguide, MIR light is generated (Fig.~\ref{fig:MIR}b). The 3.2-\textmu{}m light corresponds to $\sim 100$~THz detuning from the pump, made feasible by the higher-order dispersion present in the nanophotonic waveguide geometry (Fig.~\ref{fig:concept}(b)). While the bandwidth of the MIR light is limited by the group-velocity walkoff between the pump and signal pulses, we note that SiN waveguides can also be engineered to provide group-velocity matching and lead to broadband frequency conversion \cite{hickstein_self-organized_2018}.

Wide tunability from 2.6--3.6 \textmu{}m is observed as the waveguide widths are changed. The MIR bandwidth and power in each waveguide is approximately constant across the chip. We theoretically model the FWM/OPA process using the nonlinear Schrodinger equation \cite{agrawal_nonlinear_2012}. To calculate the dispersion, effective refractive indices are acquired via the WGModes solver \cite{fallahkhair_vector_2008} using a SiN Sellmeier equation acquired via ellipsometry measurements. The theoretical predictions agree well with the experimental measurements (Fig.~\ref{fig:MIR}(c)). The slight offset in center frequencies between theory and experiment could arise from thickness variation across the chip and waveguide. The waveguide widths on chip are fabricated in increments of 200 nm, resulting in the discrete tunability shown in Fig.~\ref{fig:MIR}c. Finer increments have been demonstrated \cite{carlson_self-referenced_2017}, and would show more continuous tunability. Light in the 2.5--2.9 \textmu{}m band shows atmospheric water absorption before detection by the FTS. 

Due to competing nonlinear processes, increasing the pump energy beyond a certain level does not increase the MIR power. To investigate this effect, we modeled the pulse propagation in a 3400-nm-wide waveguide and plot the output spectra as a function of pump-pulse energy in Fig.~\ref{fig:MIRSim}. Because the waveguides exhibit anomalous dispersion at the pump, SCG is supported in addition to FWM, leading to additional parasitic effects that limit conversion to the MIR (Fig.~\ref{fig:MIRSim}(a)).  For example, due to the long-wavelength zero-crossing in the group-velocity dispersion profiles (Fig.~\ref{fig:concept}(b)), phase-matched dispersive-wave generation in the MIR contaminates the wide-band FWM process. As seen in Fig.~\ref{fig:MIRSim}(b), at low input pump and signal pulse energies, the MIR idler grows in power linearly. However, pump pulse energies beyond 50 pJ lead to soliton fission and dispersive-wave formation that disturb the large-detuning FWM process and depletes the pump, leading to clamping of the MIR power. To mitigate the effects of SCG on the OPA process, alternative waveguide geometries such as the slot-mode waveguide could be used to provide normal dispersion at the pump while still providing phase-matching for the frequency conversion into the MIR \cite{moille_phased-locked_2018}. However, in this experiment we simply keep the pump-pulse energy below the threshold for soliton fission.

The measured MIR power is on the order of 10 \textmu W and is limited in our present devices by the weak optical confinement of the longer wavelengths and strong absorptive loss in the oxide. This linear loss for the MIR constrains the conversion efficiency in FWM \cite{agrawal_nonlinear_2012}, though we note that this may be improved by the use of suspended waveguides~\cite{halir_waveguide_2015}. Accounting for this effect and the $\sim$5 dB out-coupling loss, we estimate the on-chip power to be $\sim$100 \textmu{}W.

\begin{figure}
\centering
\includegraphics[width=\linewidth]{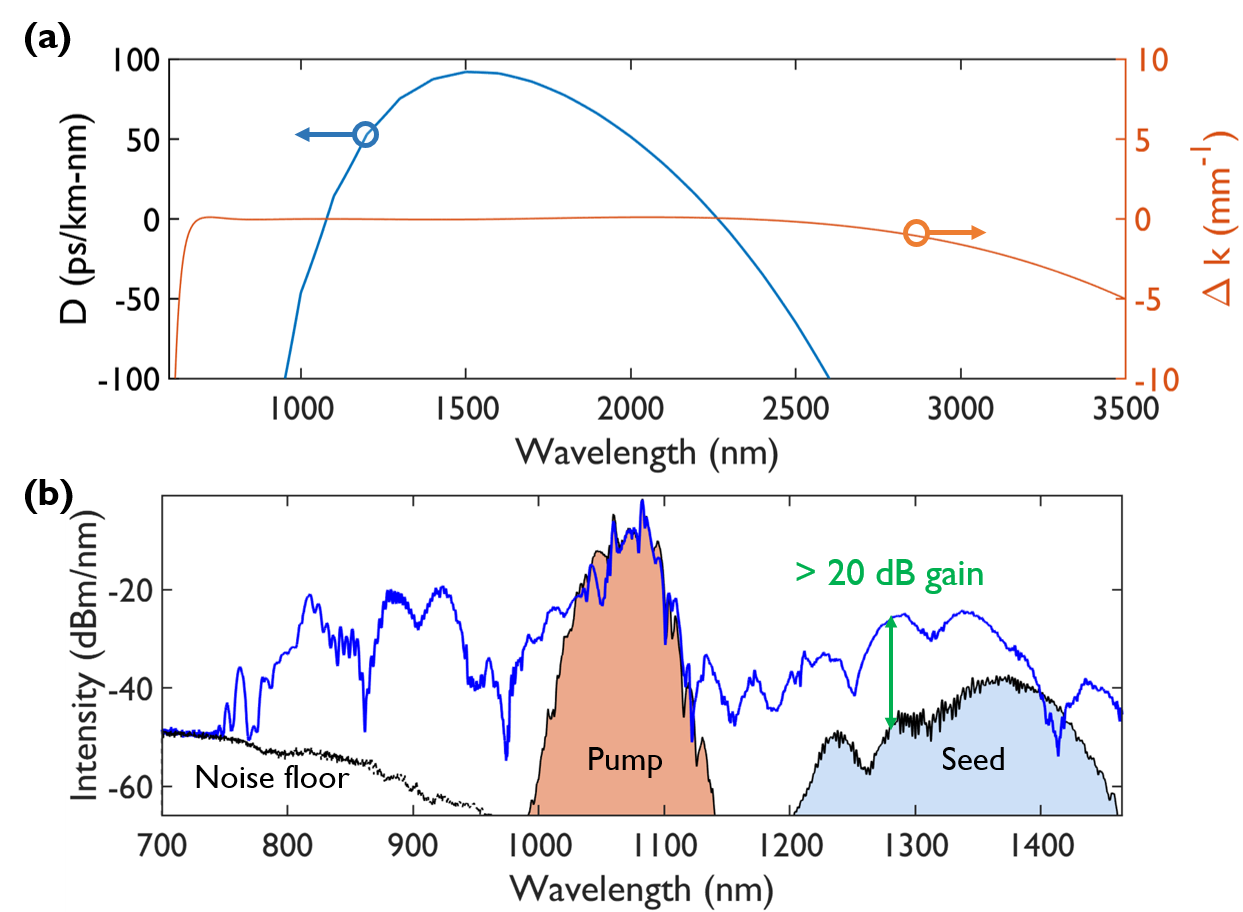}
\caption{(a) Dispersion profile (blue) for the w = 2400 nm wide waveguide on the same chip as used in Fig.~\ref{fig:MIR}, with zero-dispersion crossings at 1.07 \textmu{}m and 2.3 \textmu{}m and the corresponding FWM phase-matching curve (red)  for a pump wavelength of 1.07 \textmu{}m. The FWM phase-mismatch is near-zero for an exceptionally broad bandwidth. (b) Parametric amplification and broadband frequency conversion in the near-infrared. The 1.07 \textmu{}m pump (red, shaded) "upconverts" a 1.3~\textmu{}m seed (blue, shaded) to 0.85~\textmu{}m with high-efficiency while providing \textgreater~20~dB parametric gain (blue, solid). }
\label{fig:NIR}
\end{figure}

When strong confinement is available for all the waves, high conversion efficiency and large parametric gain can be expected in the nonlinear process. In the present waveguides, such performance is observed for an all-NIR FWM process. In order to demonstrate these advantages, we employ $\chi^{(3)}$-OPA using a 1.0-\textmu{}m (degenerate) pump and a 1.3-\textmu{}m seed in the 2400-nm-wide waveguide on the same chip. To experimentally realize these conditions, we modify the experimental setup slightly: (i) the input aspheric lens is aligned to favor coupling the 1.0\textmu m pulse (30 dB higher on-chip pulse energy than the signal) (ii) the 1.5 \textmu m pulse is spectrally broadened in HNLF and filtered, resulting in a 1.3-\textmu m pulse (50 fs pulse duration), and (iii) matched input and output aspheric lenses are used. At the pump wavelength (1.07~\textmu{}m), the group-velocity dispersion is approximately zero (Fig.~\ref{fig:NIR}(a)), allowing for an exceptionally broadband phase-matching bandwidth (Fig.~\ref{fig:NIR}(a)). When SCG is suppressed (by restricting the on-chip pump-pulse energy), net parametric gain of up to 25 dB is seen for the 1.3-\textmu{}m seed, and high-efficiency frequency upconversion into the 800-nm band is also observed (Fig.~\ref{fig:NIR}(b)). With only 300 pJ of pump-pulse energy, frequency translation exceeding 100 THz is observed along with broadband parametric gain for the seed\textemdash{}a unique combination made possible by the versatility of the integrated photonics implementation of the experiment.

In conclusion, we demonstrated tunable MIR light from 2.6--3.6 \textmu{}m in the OPA configuration using SiN nanophotonic waveguides driven by an Er:fiber laser with only 100-pJ-scale pump-pulse energies. The low pulse energy requirements make this system amenable for further photonic integration as well as GHz-repetition rate nonlinear optics \cite{carlson_ultrafast_2017}. While high absorption loss for the MIR and parasitic phase-matched processes limited the conversion efficiency, alternative waveguide designs can substantially improve MIR transmission and also impede soliton fission by providing normal dispersion at pump. Such structures can yield large, broadband parametric gain for the MIR. In this work, we demonstrated the desired performance in an all-NIR OPA implementation, showing large parametric gain and ultra broadband frequency conversion with only 300~pJ pump-pulse energy. In future work, one could integrate the spectral-broadening stage on the SiN chip and drive the self-seeded OPA process on chip using only a single laser input. This would enable MIR synthesis with a compact foot-print leveraging mature Er:fiber technology, useful for chip-scale sensing and spectroscopy applications, including quantum photonics where the spontaneously driven FWM process can yield entangled photon pairs across the NIR and MIR \cite{kalashnikov_infrared_2016}. 

\noindent DARPA  SCOUT and DODOS,  AFOSR  (FA9550-16-1-0016),  NRC, NASA, NIST.\\
\noindent The  authors  thank  Eric Stanton and Daryl Spencer for helpful comments on the manuscript. This work is a contribution of the United States government and is not subject to copyright in the United States of America.

\bibliography{Zotero}
\bibliographyfullrefs{Zotero}
 
\end{document}